\documentclass[12pt]{revtex4}
\usepackage{psfrag}
\usepackage{graphicx}
\usepackage{dcolumn}
\usepackage{bm}
\usepackage{natbib}

\begin{document}

\title{
Contact line motion for partially wetting fluids
     }

\author{
Jens Eggers
}

\affiliation{
School of Mathematics, 
University of Bristol, University Walk, \\
Bristol BS8 1TW, United Kingdom 
        }

\begin{abstract}
We study the flow close to an advancing contact line in the limit 
of small capillary number. To take into account wetting effects,
both long and short-ranged contributions to the disjoining pressure
are taken into account. In front of the contact line, there is a 
microscopic film corresponding to a minimum of the interaction potential.
We compute the parameters of the contact line solution relevant to
the matching to a macroscopic problem, for example a spreading droplet.
The result closely resembles previous results obtained with a slip 
model. 
\end{abstract}

\maketitle
\section{Introduction}
Moving contact lines are encountered in a great number of flow 
problems, such as spreading of liquid drops \cite{T79}, dewetting
of liquid films \cite{SHJ01a}, coating \cite{BR79}, and sloshing \cite{B02}.
It was
discovered by Huh and Scriven \cite{HS71} that the viscous dissipation 
in the fluid wedge bordered by a solid and a fluid-gas interface 
is logarithmically infinite if the standard hydrodynamic equations
and boundary conditions are used \cite{LL84}. Thus continuum
hydrodynamics does not describe the spreading of a drop on a table.
Instead, some microscopic length scale must be introduced into
the problem. 

As a model problem, let us consider the spreading of a viscous drop on
a flat substrate. Typical spreading speeds are so small \cite{T79}
that the bulk of the drop is almost unaffected by viscous shear
forces. Hence the drop has the shape of a spherical cap, except in
a small region around the contact line \cite{G85}. If one extrapolates
this spherical cap solution to the contact line, it meets the solid 
at a well-defined angle, called the ``apparent'' contact angle $\theta_{ap}$.
If for simplicity one assumes that the drop is thin, its radius 
$R$ is related to $\theta_{ap}$ by
\begin{equation}  
\label{ap}  
\theta_{ap} = 4V/(\pi R^3),
\end{equation}  
where $V$ is the volume of the drop.

\begin{figure}
\includegraphics[width=0.5\hsize]{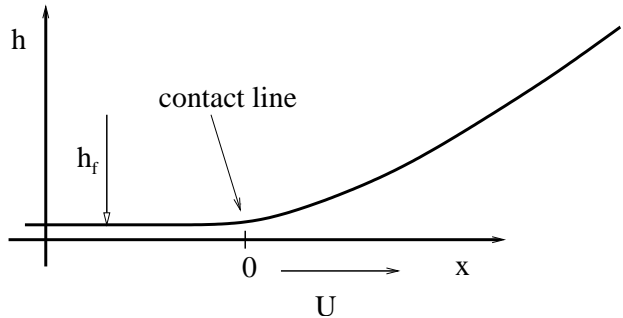}
\caption{\label{clfig} 
A cartoon of the contact line. In a frame of reference in which 
the contact line position is stationary, the solid moves to the right
with velocity $U$. There is a very thin film of thickness 
$h_f$ in front of the contact line. 
    }
\end{figure}

However, near the contact line the shear rate is of order 
$U/h$, where $U$ is the contact line speed and $h$ the local thickness
of the fluid film. Near the contact line viscous forces become very 
large, and strongly bend the interface.
A dimensionless measure of this viscous bending is the 
capillary number $Ca=\eta U/\gamma$, representing a ratio of viscous 
to capillary forces, with $\eta$ the viscosity and 
$\gamma$ surface tension. As we will show below, within the approximation
we adopt here, the slope $h'$ of the interface as function of the 
distance $x$ from the contact line has the form \cite{V76}
\begin{equation}  
\label{voinov}  
h'^3(x) = \theta_e^3 + 9Ca\ln(x/L),
\end{equation}  
where $\theta_e$ is the equilibrium contact angle and $L$ a 
microscopic length scale. As illustrated in Fig. \ref{clfig}, 
we have adopted a coordinate system in which the contact line is
at rest. The local description (\ref{voinov})
applies for $x/L\gg 1$, i.e. at a distance from the contact line
where microscopic details no longer matter. 

The distinguishing feature of (\ref{voinov}) 
is that the {\it curvature} vanishes for $x/L\rightarrow\infty$. 
This is a necessary condition for the local profile (\ref{voinov}) 
to be matchable to the spherical cap solution that makes up the bulk
of the spreading drop \cite{E05}. The details of this matching procedure
have been given in \cite{H83}, the result being
\begin{equation}  
\label{happ}  
\theta_{ap}^3 = \theta_e^3 + 9\dot{R}\eta/\gamma\ln[R/(2e^2L)],
\end{equation}  
where $e=2.718281\dots$. Together with (\ref{ap}), (\ref{happ}) is 
evidently a differential equation for the radius of the spreading drop. 
For $\theta_{ap}\gg\theta_e$ equations (\ref{ap}), (\ref{happ}) reproduce
Tanner's spreading law \cite{T79} $R = A t^{1/10}$, neglecting 
logarithmic corrections in time $t$. To find an explicit 
expression for $A$, it remains to know the length $L$. 
In this paper, we are going to compute $L$ for a model
that includes both long and short-ranged interactions in the interface
potential \cite{SHJ01a}. This model has recently become popular
for the numerical treatment of moving contact line problems
\cite{BGSMJMB03,TK03}. 

To find $L$, (\ref{voinov}) has to be continued to the contact 
line, where microscopic effects come into play. Previous calculations
\cite{H83} have done that for the case of fluid slip over the solid
surface \cite{CCSC05,KBW89}, which relieves the contact line singularity. 
In the simplest case of a Navier slip condition \cite{HS71,H83}, 
described by a slip length $\lambda$, the result is 
$L=3\lambda/(e\theta_e)$. In \cite{E04a} we have extended this 
calculation to higher orders in the capillary number. However, corrections
are found to be small in a regime where the underlying lubrication 
description is still expected to be valid \cite{CRG95}.
Apart from the slip length, an angle has to be specified at the contact
line, which is often taken to be the equilibrium contact angle. 
This assumption implies that the total dissipation near
the contact line is dominated by viscous effects, rather than dissipation
localized at the contact line \cite{BG92}. 

Here we adopt a model that builds in the equilibrium properties 
in a more rational way, by including the interface potential into
the description. Both the equilibrium contact angle \cite{I88}
and the equilibrium film thickness $h_{eq}$ are determined by 
the interface potential. Within the model, even the ``dry'' substrate
is covered by a thin film, corresponding to the minimum of the 
interface potential. The presence of this film thus formally 
eliminates the contact line singularity, $h_{eq}$ replacing the 
slip length $\lambda$ as the cut-off length. Of course, we do not 
claim that this is a true resolution of the contact line problem. 
The thickness $h_{eq}$ is often below the thickness of a single 
molecule, and even a monomolecular layer is not strictly 
describable by a continuum theory. 

Nevertheless, we believe that it is interesting to investigate the interplay
between the interface potential and viscous forces. This has first
been done by de Gennes, Hua, and Levinson \cite{GHL90}, but only taking
into account the long-ranged part of the potential. As a result, 
the equilibrium contact angle could only be worked in in an ad-hoc
fashion, as one needs the full potential to define it. We will see
below that our results are in line with the results obtained before
\cite{GHL90}. The calculation in \cite{PP04} is based on a simple 
energy balance, rather than the systematic expansion performed here. 
The very recent work \cite{PT05} treats both the advancing and the receding 
contact line in a manner very close to ours.

Our paper is organized as follows. After introducing the model
description, we recall the case of a static contact line, relating
the equilibrium contact angle to the interface potential. We then 
outline how the parameter $L$ of (\ref{voinov}) may be found in 
an expansion in the capillary number \cite{E04a}. Assuming a particular
form of the interface potential, we then solve the first order problem
explicitly. Finally, we compare to other forms of the interface 
potential as well as to previous work. 

\section{Lubrication description}
For simplicity, we perform our calculations within the framework of 
lubrication theory, thus limiting ourselves to the case of small 
contact angles, as well as small capillary number \cite{ODB97}.
Experiment shows that this approximation performs reasonably
well up to a capillary number of 0.1 \cite{CRG95}.
The lubrication equation reads \cite{BGSMJMB03}
\begin{equation}  
\label{lubd}  
3\eta \bar{h}_t=-\left[h^3(\gamma \bar{h}_{xx}+\Pi(\bar{h}))_x\right]_x,
\end{equation}
where $\bar{h}(x,t)$ is the thickness of the fluid film and 
$\Pi(\bar{h})$ is the disjoining pressure \cite{I88}. The origin
of (\ref{lubd}) is a viscous shear flow, driven by the gradient of the 
pressure $p=-\gamma \bar{h}_{xx}-\Pi(\bar{h})$. The first term is the 
usual Laplace pressure, proportional to the curvature of the interface,
while the disjoining pressure $\Pi(\bar{h})$ is given by 
$\Pi(\bar{h})=\partial V/\partial\bar{h}$, where $V(\bar{h})$ is the 
effective interface potential of a {\it flat} film of thickness 
$\bar{h}$ \cite{SHJ01a}. Thus as soon as $\bar{h}$ is larger than the range
of all the interactions between particles, $\Pi(\bar{h})$ can safely
be neglected. However, when $\bar{h}$ is of the order of a few nanometers,
the disjoining pressure becomes relevant.

To describe an
advancing contact line (cf. Fig. \ref{clfig}), it is convenient to 
pass into a frame of reference that moves with the contact 
line speed $U$:
\begin{equation}  
\label{rf}  
\bar{h}(x,t) = h(x+Ut),
\end{equation}
giving
\begin{equation}  
\label{lub}  
3 Ca h_x =-\left[h^3(h_{xx}+\Pi(h)/\gamma)_x\right]_x.
\end{equation}  

Integrating once one finds that 
\begin{equation}  
\label{lubi}  
\frac{3 Ca (h-h_f)}{h^3} = -\left[h_{xx}+\Pi(h)/\gamma\right]_x,
\end{equation}  
where $h_f$ is the (yet unknown) film thickness ahead of the 
moving contact line. 

\section{Statics}
It is instructive to look first at the well-known
static case $Ca=0$. Integrating (\ref{lubi}) once more one obtains
\begin{equation}  
\label{stat}  
P_0= -h_{xx}-\Pi(h)/\gamma,
\end{equation}  
where $P_0$ is the (constant) pressure in the film (neglecting
gravity). We are considering a situation
where the film is in contact with a large reservoir (for example a drop) 
with negligible pressure, hence $P_0=0$. Thus in the film 
we must have $\Pi(h_{eq})=0$ (corresponding to a minimum of the 
interface potential), which defines the equilibrium film thickness $h_{eq}$. 

Now (\ref{stat}) can easily be solved by putting $g(h)=h_x(x)$,
giving
\begin{equation}  
\label{stat1}  
\frac{\partial g^2}{\partial h} = -2\Pi(h)/\gamma.
\end{equation}  
Integrating (\ref{stat1}), we obtain the standard expression \cite{I88}
\begin{equation}  
\label{theq}  
\theta_e^2 = -2\int_{h_{eq}}^{\infty}\Pi(\zeta)/\gamma d\zeta
\end{equation}  
for the equilibrium contact angle, which in the lubrication 
approximation is to be identified with the slope of the
interface: $\theta_e=\tan(h_x(\infty))\approx h_x(\infty)$. By integrating
to infinity, we imply that the macroscopic scale on which $\theta_e$ 
is defined is much larger than $h_{eq}$. 

To be more specific, the disjoining pressure has a long-ranged
attractive and a short-ranged repulsive part:
\begin{equation}  
\label{dis}  
\Pi(h) = \frac{A}{6\pi h^3} - \frac{B}{h^{\alpha}}.
\end{equation}  
The repulsive interaction keeps the film thickness from collapsing 
to zero. The form of the attractive part is rather universal \cite{G85},
$A$ being known as the Hamaker constant. The most popular choice for the 
repulsive part is a power law with $\alpha=9$, which is motivated 
by the form of the Lennard-Jones interaction. Recently, enormous progress
has been made in determining the constants in (\ref{dis}) for 
some systems \cite{SHJ01a}. However, the experiments are not
sufficiently accurate to determine the value of the exponent 
$\alpha$ \cite{See05}. For some of the explicit results to be reported below we
are going to choose another value, $\alpha=5$, to be able to 
perform our calculations analytically. Using the specific form
of (\ref{dis}), one easily finds that 
\begin{equation}  
\label{eq}  
h_{eq} = (B/A)^{1/(\alpha-3)},\quad
\theta_e^2=\frac{\alpha-3}{\alpha-1}\frac{A}{6\pi\gamma h_{eq}^2}.
\end{equation}  

To compute the profile, it is useful to introduce new variables,
which are scaled to the equilibrium thickness $h_{eq}$ of the film:
\begin{equation}  
\label{scal}  
h(x) = h_{eq} H(\xi), \quad \xi=x\theta_e/h_{eq}. 
\end{equation}  
Equation (\ref{stat1}) then becomes 
\begin{equation}  
\label{stat2}  
H'^2 = 2\frac{\alpha-1}{\alpha-3}
\left(\frac{1}{H^3}-\frac{1}{H^{\alpha}}\right).
\end{equation}  
To make further progress, we specialize to $\alpha=5$, 
in which case we simply have:
\begin{equation}  
\label{statr}  
H'=\frac{H^2-1}{H^2}. 
\end{equation}  
This can be integrated to give the static interface 
shape
\begin{equation}  
\label{state}  
\xi=H+\frac{1}{2}\ln\left(\frac{H-1}{H+1}\right),
\end{equation}  
where the left hand side can of course be shifted 
by an arbitrary amount. 
\begin{figure}
\includegraphics[width=0.5\hsize]{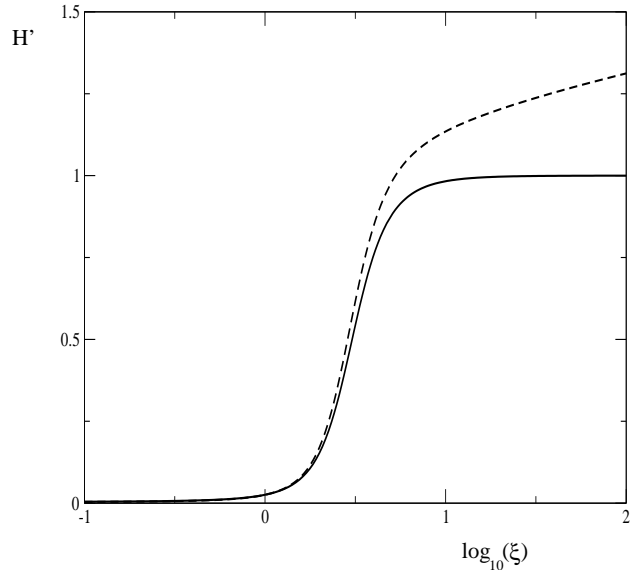}
\caption{\label{clplot} 
Numerical solutions for the rescaled interface slope $H(\xi)$. The full
line is the static solution (\ref{state}), the dashed line a 
solution of (\ref{sim}) for $\delta=0.1$ and $\alpha=5$. 
To expand the horizontal 
range, a logarithmic scale has been chosen, with $\xi=0$ shifted 
somewhat to the left of the contact line position. 
    }
\end{figure}
The slope of the static interface is shown in Fig.\ref{clplot}. To
the right of the contact line the slope asymptotes to 1, corresponding 
to the equilibrium contact angle.

\section{Perturbation expansion}
Now we turn to the problem of a moving contact line. 
In the scaled description (\ref{scal}), (\ref{lubi}) becomes
\begin{equation}  
\label{sim}  
\frac{\delta(H-H_f)}{H^3} = \left[-H''+\frac{\alpha-1}{\alpha-3}
\left(\frac{1}{H^3}-\frac{1}{H^{\alpha}}\right)\right]',
\end{equation}  
where $\delta=3Ca/\theta_e^3$ is the rescaled capillary number. 
In the limit of small $h_{eq}$ the boundaries of the system are
pushed out to $\xi=\pm\infty$, and the boundary conditions become 
\begin{equation}  
\label{bc}  
H(-\infty) = H_f, \quad H'(-\infty)=0, \quad H''(\infty)=0.
\end{equation}  
The first two conditions correspond to the assumption that 
the liquid forms a film of constant thickness ahead of the
contact line. We will see below that it {\it deviates}
slightly from the equilibrium thickness if the contact 
line is moving. The third boundary condition says 
that the curvature far away form the contact line is
vanishingly small compared to the typical curvature
near the contact line, which is $1/h_{eq}$ \cite{E04a}.

We are going to solve (\ref{sim}) in a perturbation expansion
in $\delta$, following a procedure adopted before \cite{E04a}. 
Of particular interest is the behavior of the 
solution for large $\xi$, which corresponds to (\ref{voinov}).
Namely, for $H\gg 1$ (\ref{sim}) assumes the universal form
$\delta/H^2=H'''$, which has the asymptotic solution \cite{DW97}
\begin{equation}  
\label{voix}
H'(\xi)=\left[3\delta\ln(\xi/\xi_0)\right]^{1/3}, \quad
\xi\gg 1.
\end{equation}  
This solution has vanishing curvature at infinity (as required by 
(\ref{bc})), and only contains a single free parameter $\xi_0$, 
to be determined by matching to the contact line. By comparing 
(\ref{voix}) and (\ref{voinov}), one finds 
\begin{equation}  
\label{xi0e}  
\frac{L\theta_e}{h_{eq}}=\xi_0e^{1/(3\delta)}.
\end{equation}  

On the other hand, the full solution $H(\xi)$ possesses a perturbation 
expansion in $\delta$ around the static profile $H_0(\xi)$ :
\begin{equation}  
\label{pert}  
H(\xi) = H_0(\xi) + \delta H_1(\xi) + O(\delta^2).
\end{equation}  
For large $\xi$, we have $H'_0(\xi)\approx 1$, corresponding to 
the equilibrium contact angle. By comparing this to (\ref{voix}),
we find that $\ln(\xi_0)$ has the following expansion:
\begin{equation}  
\label{xi0}  
-3\ln(\xi_0) = \frac{1}{\delta} + c1 + O(\delta).
\end{equation}  
Substituting into (\ref{voix}), we find that for large $\xi$
\begin{equation}  
\label{pertl}  
H_1'(\xi) = \ln(\xi)+c_1/3.
\end{equation}  
To compute $L$, we thus take the following steps:
First, we solve the full problem (\ref{sim}) perturbatively to
obtain $H_1(\xi)$. Then, analyzing $H_1$ for large $\xi$, we obtain $c_1$,
which gives $\xi_0$ by virtue of (\ref{xi0}). Combining this 
with (\ref{xi0e}), we finally have 
\begin{equation}  
\label{L}  
L = \frac{h_{eq}}{\theta_e}e^{-c_1/3}.
\end{equation}  

\section{Explicit solution}
To first order in $\delta$, (\ref{sim}) becomes
\begin{equation}  
\label{fo}  
\int_{-\infty}^{\xi}\frac{H_0-1}{H_0^3}d\xi=-H_1''+\frac{\alpha-1}{\alpha-3}
\left(\frac{-3H_1}{H_0^4}+\frac{\alpha H_1}{H_0^{\alpha+1}}\right) + C,
\end{equation}  
where we have integrated once, resulting in a constant of integration $C$.
From now on we consider the special case $\alpha=5$, for which we can 
make use of the static solution $H_0(\xi)$ given by (\ref{state}). 

The integral on the left-hand-side of (\ref{fo}) can be performed
by exchanging the role of dependent and independent variables
using (\ref{statr}):
\begin{equation}  
\label{foi}  
\int_{-\infty}^{\xi}\frac{H_0-1}{H_0^3}d\xi=
\int_{1}^{H_0}\frac{d H_0}{H_0(H_0+1)}dH_0=
\ln\left(\frac{2H_0}{H_0+1}\right).
\end{equation}  
The limit of (\ref{foi}) for large $\xi$ is $\ln(2)$, hence taking 
the same limit in (\ref{fo}) yields $C=\ln(2)$ for the 
constant of integration. Now considering the opposite 
limit of $\xi\rightarrow-\infty$, and using $H_0(-\infty)=1$, 
one finds $H_1(-\infty)=-\ln(2)/4$.

To solve (\ref{fo}), it is useful to rewrite 
the entire equation using $H_0$ as the independent variable.
To avoid cumbersome expressions, we denote $H_0$ by the symbol $\zeta$.
Thus (\ref{fo}) turns into:
\begin{equation}  
\label{fot}  
F(\zeta)\equiv\ln\left(\frac{\zeta}{\zeta+1}\right)=-(H_1)_{\zeta\zeta}
\left(\frac{\zeta^2-1}{\zeta^2}\right)^2+2(H_1)_{\zeta}
\left(\frac{1}{\zeta^3}-\frac{1}{\zeta^5}\right) +
6H_1\left(\frac{1}{\zeta^4}-\frac{5}{3\zeta^6}\right).
\end{equation}  
Remarkably, this equation can be solved exactly by 
noticing that two fundamental solutions are 
\begin{equation}  
\label{fund}  
H_1^{(1)}=\frac{16\zeta^5-50\zeta^3+30\zeta}{\zeta^2(\zeta^2-1)} +
15\frac{\zeta^2-1}{\zeta^2}\ln\left(\frac{\zeta-1}{\zeta+1}\right)\quad
\mbox{and} \quad H_1^{(2)}=\frac{\zeta^2-1}{\zeta^2},
\end{equation}  
which we found using Maple. Thus a general solution of (\ref{fot})
is
\begin{equation}  
\label{sol}  
H_1=H_1^{(1)}\left[b_1-\int_2^{\zeta}H_1^{(2)}F/Wd\zeta'\right] + 
H_1^{(2)}\left[b_2+\int_2^{\zeta}H_1^{(1)}F/Wd\zeta'\right],
\end{equation}  
where $W$ is the Wronskian. 

The limit $\zeta\rightarrow 1$ corresponds to the thin film.
From the condition that $H_1$ has to remain finite in this limit,
one finds
\begin{equation}  
\label{bc1}  
b_1=-\int_1^2H_1^{(2)}F/Wd\zeta'=3\ln(3)/16-\ln(2)/4,
\end{equation}  
since $H_1^{(1)}\rightarrow\infty$ for $\zeta\rightarrow 1$.
As shown in the Appendix, the other constant of integration $b_2$
is determined by the terms of order $\zeta^0$ as $\zeta\rightarrow 1$.
In the limit of $\zeta\rightarrow \infty$, on the other hand, 
one is approaching the bulk fluid, for which we find 
$H_1^{(1)}\approx16\zeta$ and $H_1^{(2)}\approx 1$, 
so a straightforward analysis of (\ref{sol}) yields
\begin{equation}  
\label{asymp1}  
H_1(\zeta)=\zeta(\ln(\zeta)-2\ln(2)) + O(\ln(\zeta)).
\end{equation}  

\section{Results and Discussion}
\begin{figure}
\includegraphics[width=0.5\hsize]{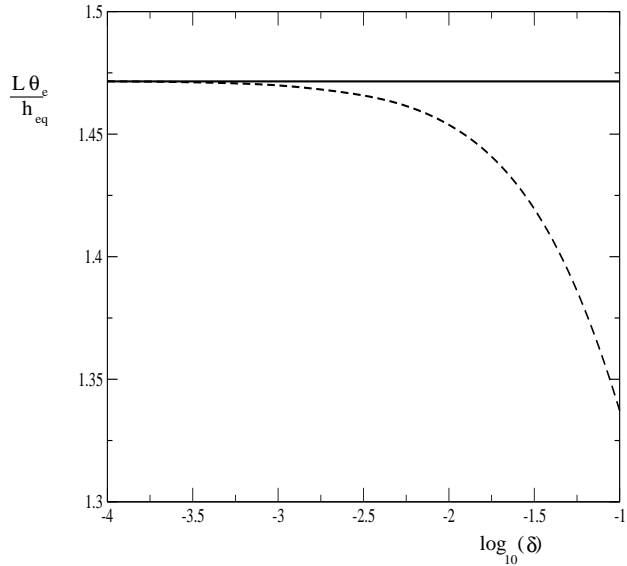}
\caption{\label{Lfig} 
A comparison with simulation. The full line is the 
leading-order result of our calculation (\ref{Lexp}), the dashed 
line is the numerical result, valid to all orders in 
$\delta$. 
    }
\end{figure}
Now we are in a position to calculate the constant $c_1$ 
appearing in (\ref{pertl}).
From (\ref{state}) we have $\zeta\equiv H_0\approx\xi$ for large $\xi$,
and thus 
\begin{equation}  
\label{asymp2}  
H_1'(\zeta)=(\ln(\xi)-2\ln(2)+1)
\end{equation}  
in this limit. We conclude that $H_1$ indeed has the asymptotic 
form (\ref{pertl}) we anticipated, and we can identify
\begin{equation}  
\label{c1}  
c_1=3-6\ln(2). 
\end{equation}  

Using (\ref{L}), we now have 
\begin{equation}  
\label{Lexp}  
L = \frac{4h_{eq}}{e\theta_e},
\end{equation}  
which is the central result of this paper. 

The result (\ref{Lexp}) can of course be tested by comparing with a 
numerical solution of the full equation (\ref{sim}). 
A linear analysis around the film thickness $H=H_f$
reveals an exponentially growing solution 
\begin{equation}  
\label{expg}  
H(\xi)=H_f+\epsilon\exp(\gamma\xi),
\end{equation}  
where $\gamma=2+O(\delta)$. Any small perturbation of 
the constant solution $H=H_f,H'=0,H''=0$ will thus lead
to an initial growth of the form (\ref{expg}). As $\xi\rightarrow\infty$,
the solution generically tends to a finite curvature \cite{DW97}. 
Thus $H_f$ has to be adjusted to find the unique solution which
obeys the boundary condition (\ref{bc}) at infinity. The asymptotics
of this solution of course has to conform with (\ref{voix}).

However, the approach to this solution is very slow, as revealed
by the full asymptotic expansion \cite{BO78}
\begin{equation}  
\label{exp}  
H'(\xi) = \left[3\delta\ln(\xi/\xi_0)\right]^{1/3}
\left\{1+\sum_{i=2}^{\infty}\frac{b_i}{(\ln(\xi/\xi_0))^i}\right\}.
\end{equation}  
To be consistent with (\ref{voix}), the coefficient $b_1$ was 
chosen to vanish, since it would lead to a redefinition of $\xi_0$.
To obtain $\xi_0$ numerically, we fitted the numerical solution 
of (\ref{sim}) to (\ref{exp}), using the first five 
terms of the expansion.
In Fig. \ref{Lfig} we plot the numerical result
for $L$ over a wide range of $\delta$-values. For reasonably
small $\delta$'s, applicable to most experimental situations,
the result is very well approximated by the present first order 
calculation. 

\begin{figure}
\includegraphics[width=0.5\hsize]{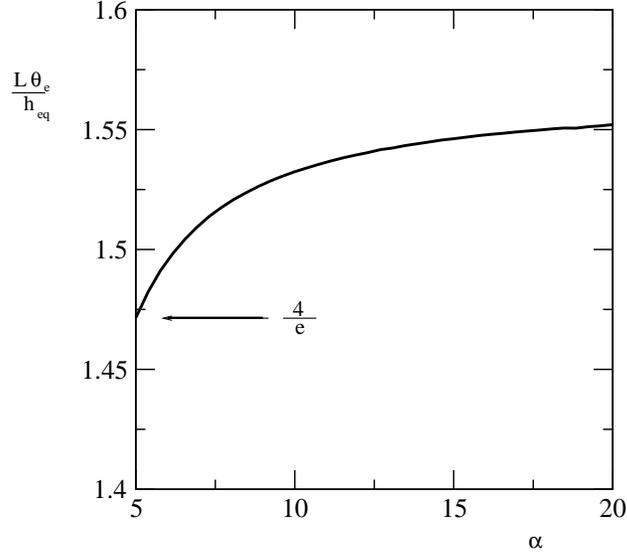}
\caption{\label{potfig} 
The characteristic length $L$ as function of the exponent
$\alpha$ characterizing the potential. For $\alpha=5$ the numerical 
result agrees with (\ref{Lexp}).
    }
\end{figure}
Our analytical approach has of course been limited to the case
$\alpha=5$, which is non-standard. Using the numerical procedure
described above, it is a simple matter to obtain $L$ for arbitrary
$\alpha$. Fig.\ref{potfig} shows the result of this calculation in the 
limit of small $\delta$. As to be expected, the variation with
$\alpha$ is not very strong. Large values of $\alpha$ correspond to
a very hard core. 

Finally, it remains to compare our results to \cite{GHL90}, who
only took the long-ranged part of the disjoining pressure into account.
At the contact line, it was assumed that the solution matches to the 
equilibrium contact angle. The result was reported in the form
$L_{GHL}=a/(2\theta_e^2)$, where 
\begin{equation}  
\label{a}  
a = \sqrt{\frac{A}{6\pi\gamma}}
\end{equation}  
is a length scale characterizing the range of van-der-Waals forces.
Thus, using (\ref{eq}) the result of \cite{GHL90} can be converted to
\begin{equation}  
\label{GHL}  
L_{GHL}=\sqrt{\frac{\alpha-1}{\alpha-3}}\frac{h_{eq}}{2\theta_e},
\end{equation}  
which is essentially the same result as (\ref{Lexp}), but with
a different prefactor. In conclusion, for both a slip and the 
present thin film model, $L$ is set by the respective microscopic
length.

\begin{acknowledgments}
I am grateful to Len Pismen for his input, and
to the participants of the Thin Film workshop in Udine in 2005, 
organized by Serafim Kalliadasis and Uwe Thiele, for advice.
\end{acknowledgments}

\appendix*
\section{}
Here we describe how to determine the remaining constant of 
integration $b_2$ in (\ref{sol}), by comparing to the asymptotics 
(\ref{expg}) of the full solution as $\xi\rightarrow-\infty$. 
Namely, as we have shown above,
\begin{equation}  
\label{expHf}  
H_f = 1 - (\ln(2)/4)\delta + O(\delta^2),
\end{equation}  
and it is straightforward to see that the exponent is 
\begin{equation}  
\label{expgamma}  
\gamma = 2 + \gamma_1\delta + O(\delta^2), \quad \gamma_1=9\ln(2)/4-1/8.
\end{equation}  
Thus at zeroth order in $\delta$ one finds $\zeta=1+\epsilon\exp(2\xi)$.
On the other hand, the full static profile (\ref{state}) gives 
$2(\xi-1+\ln(2)/2) = \ln(\zeta-1) +O(\zeta-1)$. Thus by comparing 
the two profiles one identifies $\ln(\epsilon)=\ln(2)-2$.

Expanding (\ref{expg}) to next order in $\delta$ leads to 
\begin{eqnarray}  
\label{exph1}  
H_1=-\ln(2)/4+\gamma_1(\xi-1+\ln(2)/2)\exp[2\xi-2+\ln(2)]= \\
-\ln(2)/4+\gamma_1\ln(\zeta-1)(\zeta-1) + O(\zeta-1)^2. \nonumber
\end{eqnarray}  
Thus in the limit of $\zeta\rightarrow 1$, (\ref{sol}) must have 
the same form as (\ref{exph1}). The integrals in (\ref{sol}) can 
be performed using Maple, and in the limit they give 
\[
H_1 = -\ln(2)/4 + \left(a + \gamma_1\ln(\zeta-1)\right)(\zeta-1),
\]
which matches (\ref{exph1}) if $a=0$. From this requirement 
we finally get
\begin{eqnarray}  
\label{b2}  
b_2={\frac {107}{192}}-{\frac {157}{96}}\,\ln  \left( 3 \right) -
3\,{\it dilog} \left( 2/3 \right) +3/2\,{\it dilog} \left( 4/3 \right) + \\
{\frac {15}{16}}\,{\it dilog} \left( 3 \right) + {\frac {75}{32}}\,
 \left( \ln  \left( 3 \right)  \right) ^{2}+{\frac {125}{48}}\,\ln 
 \left( 2 \right) - \nonumber \\
{\frac {27}{8}}\,\ln  \left( 2 \right) \ln  \left( 
3 \right) -{\frac {15}{16}}\, \left( \ln  \left( 2 \right)  \right) ^{
2}+{\frac {13}{32}}\,{\pi }^{2} = 0.359777\dots. \nonumber
\end{eqnarray}  

\bibliography{paper2}

\end{document}